\documentclass[runningheads]{llncs}
\usepackage{graphicx}
\usepackage{adjustbox}
\usepackage{eucal}
\usepackage{mathrsfs}
\usepackage{tikz}
\usepackage{soul}
\usepackage{pgfplots}
\usepackage{hyphenat}
\usepackage{pgfplotstable}
\usepackage{comment}
\usetikzlibrary{calc}
\usetikzlibrary{shapes}
\usepackage{bm}
\usepackage{caption}
\usepackage{multirow}
\usepackage{authblk}

\usepackage{amsmath,amsfonts,amssymb,epsfig,epstopdf,url,array}

\definecolor{cadmiumgreen}{rgb}{0.0, 0.42, 0.24}
\definecolor{darkolivegreen}{rgb}{0.33, 0.42, 0.18}
\title{Graph Neural Network based scheduling : Improved throughput under a generalized interference model \thanks{The authors are grateful to the OPAL infrastructure from Université Côte d'Azur for providing resources and support. This paper is accepted in EAI VALUETOOLS 2021. This work was funded by ANR MASTERO 5G.}}
\titlerunning{Graph Neural Network based scheduling}
\vspace{-3em}
\date{}
\begin{document}

\begin{comment}
\author{First Author\inst{1}\orcidID{0000-1111-2222-3333} \and
Second Author\inst{2,3}\orcidID{1111-2222-3333-4444} \and
Third Author\inst{3}\orcidID{2222--3333-4444-5555}}
%
\authorrunning{F. Author et al.}
\institute{Princeton University, Princeton NJ 08544, USA \and
Springer Heidelberg, Tiergartenstr. 17, 69121 Heidelberg, Germany
\email{lncs@springer.com}\\
\url{http://www.springer.com/gp/computer-science/lncs} \and
ABC Institute, Rupert-Karls-University Heidelberg, Heidelberg, Germany\\
\email{\{abc,lncs\}@uni-heidelberg.de}}
%
\maketitle  
\end{comment}
%\begin{comment}
\author{S. Ramakrishnan \inst{1} (ramakrishnan.sambamoorthy@inria.fr) \and
Jaswanthi Mandalapu\inst{2} (ee19d700@smail.iitm.ac.in)\and
Subrahmanya Swamy Peruru\inst{3} (swamyp@iitk.ac.in) \and 
Bhavesh Jain\inst{3} (jbhavesh@iitk.ac.in)\and
Eitan Altman\inst{1,2} (eitan.altman@inria.fr)} 
\vspace{-1.5em}
\authorrunning{S. Ramakrishnan et al.}
\institute{INRIA Sophia Antipolis - M\'editeran\'ee, and Univ of Avignon, France; \and LINCS Lab, paris; \and
IIT Madras, Chennai, India;  \and 
IIT Kanpur, Kanpur, India. }
%\email{\{{jbhavesh, swamyp}@iitk.ac.in\}}
%}
%
\maketitle 
%\end{comment}
% typeset the header of the contribution
%
\vspace{-2.5em}
\begin{abstract}
In this work, we propose a Graph Convolutional Neural Networks (GCN) 
based scheduling algorithm for adhoc networks. In particular, we 
consider a generalized interference model called the $k$-tolerant 
conflict graph model and design an efficient approximation for the 
well-known Max-Weight scheduling algorithm. A notable feature of 
this work is that the proposed method do not require labelled data set (NP-hard to compute) for training the neural
network.  Instead, we design a loss function that utilises the 
existing greedy approaches and trains a GCN that improves the 
performance of greedy approaches. 
Our extensive numerical experiments illustrate
that using our GCN approach, we can  
significantly ($4$-$20$ percent) improve the performance of the conventional greedy approach. 

\end{abstract}

\vspace{-2em}
\section{Introduction}
The design of efficient scheduling algorithms is a fundamental problem in wireless networks. In each time slot, a scheduling algorithm aims to determine a subset of non-interfering links such that the system of queues in the network is stabilized. Depending on the interference model and the network topology, it is known that there exists a \textit{`rate region'} - a maximal set of arrival rates - for which the network can be stabilized. A scheduling algorithm that can support any arrival rate in the rate region is said to be throughput optimal. A well-known algorithm called the Max-Weight scheduling algorithm \cite{182479} is said to be throughput optimal. However, the Max-Weight scheduler is not practical for distributed implementation due to the following reasons: (i) global network state information is required, and (ii) requires the computation of maximum-weighted independent set problem in each time slot, which is an NP-hard problem.

There have been several efforts in the literature to design low-complex, distributed approximations to the Max-Weight algorithm \cite{11222334,max_w_2}. Greedy approximation algorithms such as the \textit{maximal} scheduling policies, which can support a fraction of the maximum throughput, are one such class of approximations \cite{Wan2013}. On the other hand, we have algorithms like carrier sense multiple access (CSMA) algorithms \cite{csma_1,csma_2}, which are known to be near-optimal in terms of the throughput performance but known to suffer from poor delay performance.

Inspired by the success of deep-learning-based algorithms in various fields like image processing and natural language processing, recently, there has been a growing interest in their application in wireless scheduling as well \cite{ml_1,ml_2,ml_3}. Initial research in this direction focused on the adaption of widely used neural architectures like multi-layer perceptrons or convolutional neural networks (CNNs) \cite{cnn_1} to solve wireless scheduling problems. However, these architectures are not well-suited for the scheduling problem because they do not explicitly consider the network graph topology. Hence, some of the recent works in wireless networks study the application of the Graph Neural Network (GNN) architectures for solving the scheduling problem \cite{gnn_1}. For instance, a recent work \cite{9414098} has proposed a GNN based algorithm, where it has been observed that the help of Graph Neural networks can improve the performance of simple greedy scheduling algorithms like Longest-Queue-First (LQF) scheduling.

However, this result is observed on a simple interference model called the conflict graph model, which captures only binary relationships between links. Nevertheless, in real wireless networks, the interference among the links is additive, and the cumulative effect of all the interfering links decides the feasibility of any transmission. Hence, it is essential to study whether the GNN based approach will improve the performance of greedy LQF scheduling under a realistic interference model like the (Signal-to-interference-plus-noise ratio) SINR model, which captures the cumulative nature of interference.

One of the challenges in conducting such a study is that the concept of graph neural networks is not readily applicable for the SINR interference model since a graph cannot represent it. Hence, we introduce a new interference model which retains the cumulative interference nature yet is amenable to a graph-based representation and conduct our study on the proposed interference model. This approach will provide insights into whether the GNN-based improvement for LQF will work for practical interference models.

To that end, in this paper, we study whether GNN based algorithms can be used for designing efficient scheduling under this general interference model. Specifically, we consider a $k$-tolerant conflict graph model, where a node can successfully transmit during a time slot if not more than $k$ of its neighbors are transmitting in that time slot. Moreover, when $k$ is set to zero, the $k$-tolerance model can be reduced to the standard conflict graph model, in which a node cannot transmit if any of its neighbors is transmitting. We finally tabulate our results and compare them with other GNN-based distributed scheduling algorithms under a standard conflict-graph-based interference model. 
In sum, our contributions are as follows:
\begin{itemize}
\item[(i)] We propose a GCN-based distributed scheduling algorithm for a generalized interference model called the $k$-tolerant conflict graph model. 
\item[(ii)] The training of the proposed GCN does 
not require a labeled data set (involves solving an NP-hard 
problem). Instead, we design a loss function that utilizes an existing greedy approach and trains a GCN that improves the performance of the greedy approach by $4$ to $20$ percent.
\end{itemize}
The remainder of the paper is organized as follows. In Sec.~\ref{section2}, we briefly present our network model. In Sec.~\ref{section3}, an optimal scheduling policy for $k$-tolerance conflict graph interference model, a GCN-based $k$-tolerant independent set solver, is presented. In Sec.~\ref{section4}, we conduct experiments on different data sets and show the numerical results of the GCN-based scheduling approach. Finally, the paper is concluded in Sec.~\ref{section5}.\\

\noindent
\textbf{\textit{Motivation:}} In the SINR interference model, a link can successfully transmit if the cumulative interference from all nodes within a radius is less than some fixed threshold value. The conflict graph model insists that all the neighbours should not transmit when a link is transmitting. However, in a real-world situation, a link can successfully transmit as long as the cumulative interference from all its neighbours (the links which can potentially interfere with a given link) is less than a threshold value. As a special case, in this paper, we consider a conservative SINR model called k-tolerance model in which, if $i_{max}$ is the estimated strongest interference that a link can cause to another and let  $i_{th}$ be the cumulative threshold interference that a link can tolerate, then a conservative estimate of how many neighbouring links can be allowed to transmit without violating the threshold interference is given by $k= i_{th}/i_{max}$. In other words, $k$-neighbours can transmit while a given link is transmitting. %This is referred to as the $k$-tolerance interference model. 
It can be seen that this conservative model retains the cumulative nature of the SINR interference model. Hence a study on this model should give us insights into the applicability of GNN based solutions for realistic interference models.

\section{Network Model}\label{section2}
We model the wireless network as an undirected graph 
$\mathcal{G} = (V,E)$ with $N$ nodes. 
Here, the set of nodes $V = \{v_i\}_{i=1}^N$ of the graph 
represents links in the wireless network i.e., 
a transmitter-receiver pair. 
We assume an edge between two nodes, if 
the corresponding links could potentially interfere 
with each other.
Let $E$ and $\textbf{A}$ denote the set of edges and the adjacency matrix of 
graph $\mathcal{G}$ respectively. We denote the set 
of neighbors of node $v$ by $\mathcal{N}(v)$ i.e.,
%For any node $v \in V$, 
%let $\mathcal{N}(v)$ defines the set of neighbors of $v$. 
a node $v^{\prime} \in \mathcal{N}(v)$, 
if the nodes $v$ and $v^{\prime}$ share an edge between them.  
We say a node is $k$-tolerant,
if it can tolerate at most $k$ of its transmitting neighbors. In other words, a 
$k$-tolerant node can successfully transmit, if the number of neighbors 
transmitting at the same time is at most $k$.  
We define a \textit{$k$}-\textit{tolerant conflict graph} as a graph in which each node 
is $k$-tolerant, and model the wireless network as a 
\textit{$k-$tolerant conflict graph}. Note that this is a generalization of the 
popular conflict graph model, where a node can tolerate none of 
its transmitting neighbors. The conflict graph model corresponds to $0$-tolerant conflict
graph ($k=0$).

We assume that the time is slotted. In each time slot, 
the scheduler has to decide on the set of links to transmit in that time slot.
A feasible schedule is a set of links that can successfully transmit at the same time.
At any given time $t$, a set of links can successfully transmit, 
if the corresponding nodes form a \textit{$k$-independent set} (defined below) in graph
$\mathcal{G}$. 
Thus, a feasible schedule corresponds to a $k$-independent set in $\mathcal{G}$.

\begin{definition}
($k$-independent set) A subset of vertices of a 
graph~$\mathcal{G}$ is $k$\hyp{}independent, if it induces 
in $\mathcal{G}$, a sub-graph of maximum degree at most $k$.
\end{definition}
\par A scheduler has to choose a feasible schedule at any given time.
Let $\mathcal{S_G}$ denotes the collection of all possible $k-$independent sets i.e., the 
feasible schedules. We denote the schedule at time $t$ by an $N$ 
length vector $\sigma(t) =\left(\sigma_v(t), \; v \in V\right)$. 
We say $\sigma_v(t) = 1$ if at time $t$, node $v$ is scheduled to transmit and 
$\sigma_v(t)=0$, otherwise.
Depending on the scheduling decision $\sigma(t) \in \mathcal{S_G}$
taken at time $t$, node $v \in V$ (a link in the original wireless network) 
gets a rate of $\mu_v(t,\sigma)$. 
We assume that packets arriving at node $v$ can be stored in an infinite buffer.  
At time $t$, let $\lambda_v(t)$ be the number of packets that arrive at node 
$v \in V$. We then have the following queuing dynamics at node $v$:
\begin{align}
q_v(t+1) = \left[q_v(t) + \lambda_v(t) - \mu_v(t,\sigma) \right]^+ .   
\end{align}
The set of arrival rates for which there exist a scheduler that 
can keep the queues stable is known as the rate region 
of the wireless network.

\subsection{Max-Weight Scheduler}
A well known scheduler that stabilises the network is the Max-Weight algorithm~\cite{182479}.
The Max-Weight algorithm chooses a schedule $\sigma^*(t) \in \mathcal{S_G}$ that 
maximizes the sum of queue length times the service rate, i.e., 
\begin{align}
\label{eqn: Max-Weight algorithm}
    \sigma^*(t) = \arg \max_{\sigma \in \mathcal{S_G}} \sum_v q_v(t) 
    \mu_v(t,\sigma).
\end{align}
We state below one of the celebrated results in radio resource allocation.
\begin{theorem} \cite{182479}
Let the arrival process $\lambda_v(t)$ be an 
ergodic process with mean $\lambda_v$.
If the mean arrival rates ($\lambda_v$) are within the rate 
region, then the Max-Weight 
scheduling algorithm is throughput optimal.
\end{theorem}
In spite of such an attractive result, the Max-Weight algorithm is seldom implemented in practice. This is because, the scheduling decision in \eqref{eqn: Max-Weight algorithm} has complexity that is exponential in the number of nodes. 
Even with the simplistic assumption of a conflict graph model, 
\eqref{eqn: Max-Weight algorithm} reduces to the NP-hard problem of finding the maximum weighted independent set. At the timescale of these scheduling decisions, finding the exact solution to \eqref{eqn: Max-Weight algorithm} is practically infeasible. Hence, we resort to solving \eqref{eqn: Max-Weight algorithm} using a Graph Neural Network (GNN) model. Before we explain our GNN based algorithm, we shall rephrase the problem in \eqref{eqn: Max-Weight algorithm} for the k-tolerant conflict graph model below.

\subsection{Maximum weighted k-independent set}
In the $k$-tolerant conflict graph model $\mathcal{G}$, the Max-Weight problem 
is equivalent to the following integer program:
\begin{align}
%\left.
\label{eqn:k-independent set problem}
    \begin{aligned}
        \mbox{Maximize: } & \sum_v \sigma_v w_v \\
        \mbox{Such that: } & \sigma_v  \left(\sum_{v^{\prime} \in \mathcal{N}(v)} 
        \sigma_{v^\prime}\right) \leq  k \\
         & \sigma_{v} \in \{0,1\}, \text{ for all } v \in \mathcal{V}
    \end{aligned}
%\right \}
\end{align}
Here $\bm{w} = (w_v:\; v \in V)$ is the weight vector. The constraint in 
\eqref{eqn:k-independent set problem} ensures that whenever a node is 
transmitting, at most $k$ of its neighbors can
transmit.
It can be observed that the maximum weight 
problem in \eqref{eqn: Max-Weight algorithm}
corresponds to using the weights 
$w_v = q_v(t) \mu_v(t,\sigma)$ in the above formulation. 
Henceforth, the rest of this paper is devoted to solving the maximum weighted 
$k$-independent set problem using a graph neural network.\\
\section{Graph Neural Network based Scheduler}\label{section3}
\begin{center}
\begin{figure}[h!]
    \hspace{-1.1in}
    \includegraphics[width = 1.5\textwidth]{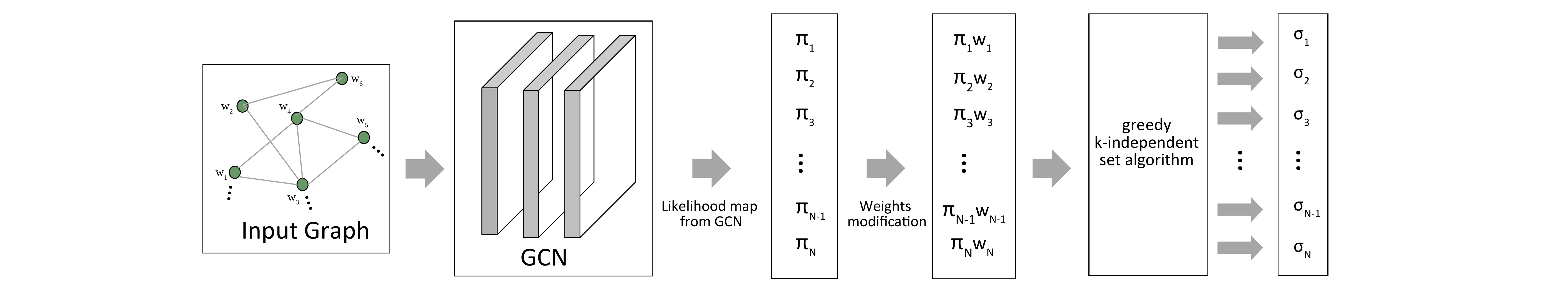}
    \caption{The architecture of the Graph Convolutional Neural Network based maximum weighted $k-$independent set problem solver.}
    \vspace{-2em}
    \label{fig:gcn_schematic}
\end{figure}
\end{center}
In this section, we present a graph neural network 
based solution to solve the maximum weighted 
$k$-independent set problem. We use the Graph Convolution 
Neural network (GCN) architecture from \cite{kipf2017semi,graph_conv}. 

The GCN architecture is as follows:
We use a GCN with $L$ layers. The input of each layer is a 
feature matrix $\textbf{Z}^l \in \mathbb{R}^{N \times C^l} $ 
and its output is fed as the input to the next layer. 
Precisely, at the $(l+1)$th layer, the feature matrix $\textbf{Z}^{l+1}$ is
computed using the following graph convolution operation:
\begin{align}
  \textbf{Z}^{l+1} = \Phi(\textbf{Z}^l \bm{\Theta}_0^l + \bm{\mathcal{L}} \textbf{Z}^l \bm{\Theta}_1^l),
\end{align}
where $\bm{\Theta}_0^l, \bm{\Theta}_1^l \in \mathbb{R}^{C^l \times C^{l+1}}$  are the trainable weights of the neural network, $C^l$~denotes the number of feature channels 
in $l$-th layer, $\Phi(.)$ is a nonlinear activation function and $\bm{\mathcal{L}}$ is the normalized Laplacian of the input graph $\mathcal{G}$ computed as follows:
%\begin{align*}
    $\bm{\mathcal{L}} = \textbf{I}_N - 
    \textbf{D}^{-\frac{1}{2}} \textbf{A} \textbf{D}^{-\frac{1}{2}}.$
%\end{align*}
Here, $\textbf{I}_N$ denotes the $N \times N$ identity matrix and $\textbf{D}$ is the diagonal matrix with entries $\textbf{D}_{ii} = \sum_j \textbf{A}_{ij}$.

\par We take the input feature matrix 
$\textbf{Z}^0 \in \mathbb{R}^{N \times 1}$ 
as the weights $\bm{w}$ of the nodes (hence $C^0 = 1$) 
and $\Phi(.)$ as a ReLU activation function 
for all layers except for the last layer. For the last layer, 
we apply sigmoid activation
function to get the likelihood of the nodes to be included in the $k$-independent set. 
We represent this likelihood map from the GCN network using an 
N length vector $\bm{\pi} = (\pi_v,\; v \in V) \in [0,1]^N$. 

In summary, the GCN takes a graph $\mathcal{G}$ and the node 
weights $\bm{w}$ as input and returns a $N$ length 
likelihood vector $\bm{\pi}$ (see Figure \ref{fig:gcn_schematic}).
However, we need a $k$-independent 
set. In usual classification problems, 
such a requirement is satisfied by projecting the likelihood 
maps to a binary vector. Projecting the likelihood map onto the 
collection of $k$-independent sets is not straightforward, since 
the collection of $k$-independent sets are $N$ length binary vectors 
that satisfy the constraints in \eqref{eqn:k-independent set problem}.
Such a projection operation by itself might be costly in terms of 
computation. Instead, by taking inspiration from 
\cite{NEURIPS2018_8d3bba74}, we pass the likelihood map through 
a greedy algorithm\footnote{In practice, the greedy algorithm can be replaced with a distributed greedy algorithm \cite{7084695} and train the GCN model w.r.t the distributed greedy algorithm.} to get a $k$-independent set. 

The greedy algorithm requires each node to keep track of the 
number of its neighbours already added in $k$-independent set. 
We sort the nodes in the descending order of the product of the 
likelihood and the weight i.e., $\pi_v w_v$. We add the   
node with highest likelihood-weight product to the 
$k$-independent set, if at most $k$ of its neighbors 
are already added in the $k$-independent set. 
We remove the nodes that are neighbours to a node 
which has already added to the set and also reached a tolerance 
of $k$. We then repeat the procedure until 
no further nodes are left to be added.

We use a set of node-weighted graphs to train the GCN. Since 
the problem at hand is NP-hard, we refrain from finding 
the true labels (maximum weighted $k$-independent set) to train the 
GCN. Instead, we construct penalty
and reward functions using the desirable properties of the output $\bm{\pi}$. We then
learn the parameters by optimizing over a weighted sum of the constructed
penalties and rewards. We desire the output
$\bm{\pi}$
to predict the maximum weighted
$k$-independent set. 
With this in mind we construct the 
following rewards and penalties:

\par  
%Thus, 
%we design the following components of the cost 
%function assuming that $\bm{\pi}$ represents 
%the solution of the maximum weighted $k$-independent set problem.

\begin{enumerate}
    \item [a)] The prediction $\bm{\pi}$ needs to maximize the sum of the weights. So, our prediction needs to maximize
    %\begin{align*}
     $ R_1 = \sum_v \pi_v w_v $.
    %\end{align*}
    \item [b)] The prediction $
    \bm{\pi}$ needs 
    to satisfy the 
    $k$-independent set constraints. 
    Therefore, we add a penalty, 
    if  $\bm{\pi}$ violates the independent set 
    constraints in \eqref{eqn:k-independent set problem}, i.e.,
    %\begin{align*}
        $P_1 = \sum_{v \in V} \left(\sigma_v \Big( \sum_{v^{\prime} \in \mathcal{N}(v)} 
        \sigma_{v^\prime} - k \Big) \right)^2$.
    %\end{align*}
    \item [c)] Recall that we use the greedy algorithm to predict the $k$-independent set from $\bm{\pi}$. The 
    greedy algorithm takes $(\pi_v w_v, \; v \in V)$ as the 
    input and returns a $k$-independent set.
    We desire the total weight of the output 
    $\bm{\pi}$, i.e., $\sum_v \pi_v w_v$ to be close to the 
    total weight of the $k$-independent returned by the 
    greedy algorithm. Let $W_{gcn}$ be the total weight of 
    the independent set predicted by the 
    greedy algorithm. Then, we penalise the 
    output $\bm{\pi}$ if it deviates from $W_{gcn}$, i.e.,
    %\begin{align*}\label{eq_cost_function}
        $P_2 = \left|\sum_v \pi_v w_v - W_{gcn} \right|^2$.
    %\end{align*}
\end{enumerate}

We finally construct our cost function as a weighted sum of the 
above i.e., we want the GCN to minimize the cost function:
\begin{align}\label{eq_cost_function}
    C = \beta_1 P_1 + \beta_2 P_2 - \beta_3 R_1
\end{align}
where $\beta_1$, $\beta_2$ and $\beta_3$ denotes the optimization weights of the cost function defined in equation~\eqref{eq_cost_function}.
%\pgfplotstableread{./Results/K0_ER.csv} 
%\pgfplotstableread{./Results/K0_BA.csv} 
\section{Experiments}\label{section4}
We perform our experiments on a single GPU GeForce GTX 1080 Ti
\footnote{Training the models took around two hours.}.
The data used for training, validation and testing 
are described in the subsection below.
\subsection{Dataset}
\par We train our GCN using randomly generated graphs.
We consider two graph distributions, namely Erdos-Reyni (ER) and 
Barbasi-Albert (BA) models. These distributions were also used in \cite{9414098}. 
Our choice of these graph models is to ensure fair comparison with prior work 
on conflict graph model \cite{9414098} ($k=0$). 

In ER model with $N$ nodes, an edge is introduced between two nodes with 
a fixed probability $p$, independent of the generation of other edges. 
The BA model generates a graph with $N$ nodes (one node at a time),
preferentially attaching the node to $M$ existing nodes with probability 
proportional to the degree of the existing nodes. 
\par For training purpose, we generate $5000$ graphs of each of 
these models. For the ER model, we choose 
$p \in \{0.02, 0.05, 0.075, 0.10, 0.15\}$ and for the 
BA model we choose $M = Np$. The weights of the nodes are chosen 
uniformly at random from the interval $[0,1]$. We use an additional 
$50$ graphs for validation and $500$ graphs for testing.

\subsection{Choice of hyper-parameters}
We train a GCN with $3$ layers consisting i) an input layer 
with the weights of the nodes as input features
ii) a single hidden layer with $32$ features and 
iii) an output layer with $N$ features (one for each node)
indicating the likelihood of choosing the corresponding node in
the $k$-independent set. 
This choice of using a smaller number of layers ensures 
that the GCN operates with a minimal number of communications 
with its neighbors. We fix $k = 0$, and experiment training the GCN 
with different choices of the optimization weights $\beta_1$, $\beta_2$
and $\beta_3$. The results obtained are tabulated in Figure \ref{fig:table1}. Let $W_{gr}$ denote the total weight of 
the plain greedy algorithm i.e., without any GCN and $W_{gcn}$ 
denote the total weight of the independent set predicted by the
GCN-greedy combination. We have tabulated the average ratio 
between the total weight of the nodes in the independent set 
obtained from the GCN-greedy and the total weight 
of the nodes in the independent set obtained from the plain greedy algorithm, i.e., $W_{gcn}/W_{gr}$. The average is taken over the 
test data set.
\begin{figure}
\begin{center}
\begin{tabular}{ |c|c|c|c|c|c|c|c| } 
\hline
 & & & & \multicolumn{2}{c|}{Test Data = ER} &  
 \multicolumn{2}{c|}{Test Data = BA}  \\ \cline{5-8} 
Training Data & $\beta_1$ & $\beta_2$ & $\beta_3$ & Average & Variance & Average & Variance \\
& & & &$W_{gcn}/W_{gr}$ & $\times \; 10^{-3}$ & $W_{gcn}/W_{gr}$& $\times \; 10^{-3}$  \\
\hline
\multirow{8}{*}{BA} &	5	&	5	&	10	&	1.038 	&	3.047
	&	1.11 & 10.16	\\
&	10	&	10	&	1	&   1.035 	&	3.297 &	 1.11	&	10.37	\\
&	5	&	5	&	1	&	1.035 	&	3.290 &	1.11	&	10.14	\\
&	1	&	1	&	1	&	1.034 	&	3.253	&	1.10	&	10.23	\\
&	5	&	5	&	30	&	1.041 	&	3.230	&	1.10	&	10.39	\\
&	5	&	5	&	50	&	1.041 	&	3.214 &	1.10	&	10.28	\\
&	5	&	5	&	100	&	1.035 	&	2.838 &	1.09	&	10.02	\\
&	30	&	1	&	1	&	1.031 	&	2.401 &	1.07	&	8.25	\\
 \hline
\multirow{8}{*}{ER} & 5 & 5 & 30 & 1.040 & 2.929
& 1.10	&	10.12 \\
 & 5 & 5 & 10 & 1.039 & 3.145 &
1.11	&	10.71	\\
 & 5 & 5 & 50 & 1.039 & 2.957 &
1.09	&	9.92 \\ 
 &	1	&	1	&	1	&	1.038 	&	3.135	&	1.11	&	10.74 \\
&	1	&	20	&	1	&	1.036 	&	3.070	&	1.11	&	10.55 \\
&	10	&	10	&	1	&	1.034 	&	3.428	&	1.11	&	10.34 \\
&	5	&	5	&	1	&	1.034 	&	3.331	&	1.11	&	10.34\\
&	5	&	5	&	100	&	1.031 	&	2.420	&	1.08	&	8.42\\
 \hline
 \multicolumn{4}{|c|}{Distributed scheduling using GNN 
 \cite{9414098}} & 1.039 & 3.5 & 1.11 & 11.0 \\
 \hline
\end{tabular}
\caption{Table showing the average and variance of the ratio of the total 
weight of the nodes in the independent set ($K=0$) obtained using GCN to 
that of the independent set obtained using greedy algorithm. 
We observe a $3$ percent increase in the total weight for the ER model and 
$11$ percent increase in the total weight for the BA model. 
Our performance matches with the performance of the GCN used in
\cite{9414098}.}
\vspace{-1.5em}
\label{fig:table1}
\end{center}
\end{figure}
The training was done with BA and ER models separately. We test 
the trained models also with test data from both models to understand if the trained models are transferable. 
We see that GCN trained with  
parameters $\beta_1 = 5$, $\beta_2 = 5$ and $\beta_3 = 10$
performs well 
for both ER and BA graph models. 
The GCN improves the total weight of the 
greedy algorithm by $4$ percent for the ER model and 
by $11$ percent for the BA model.
Also, we see that the GCN trained with ER
model performs well with BA data and vice versa.  

\subsection{Performance for different $k$}
We also evaluate the performance for different tolerance values
$k \in \{1,2,3,4\}$. We use the parameters 
$\beta_1 = 5$, $\beta_2 = 5$ and $\beta_3 = 10$ 
in the cost function. Recall that we have come up with this 
choice using extensive simulations for $k=0$.
In Figure~\ref{fig:table2}, we tabulate the average ratio 
between the total weight of the $k$-independent set obtained 
using the GCN-greedy combo and that of the plain greedy algorithm
i.e., $W_{gcn}/W_{gr}$. 
We have also included the variance from this performance. 
We observe that the performance for a general $k$ is even better 
as compared to $k=0$. For example, we see that for $k = 2,3,4$, 
we see  $6$ percent improvement for the ER model and close to $20$
percent improvement for the BA model.
\begin{figure}
\begin{center}
\begin{tabular}{ |c|c|c|c|c|c| } 
\hline
 & & \multicolumn{2}{c|}{Test Data = ER} &  
 \multicolumn{2}{c|}{Test Data = BA}  \\ \cline{3-6} 
Training Data & $k$ & Average & Variance & Average & Variance \\
& &$W_{gcn}/W_{gr}$ & $\times \; 10^{-3}$ & $W_{gcn}/W_{gr}$& $\times \; 10^{-3}$  \\
\hline
    \multirow{4}{*}{BA} &	1	&	1.056	&	4.07	&	1.143	&	10.22	\\
&	2	&	1.062	&	5.26	&	1.193	&	10.92	\\
&	3	&	1.067	&	5.55	&	1.209	&	20.14	\\
&	4	&	1.063	&	4.53	&	1.241	&	20.57	\\
\hline
\multirow{4}{*}{ER} &	1	&	1.056	&	3.99	&	1.143	&	10.18	\\
&	2	&	1.064	&	5.12	&	1.187	&	10.81	\\
&	3	&	1.066	&	4.82	&	1.205	&	20.13	\\
&	4	&	1.062	&	4.18	&	1.225	&	20.29	\\
 \hline
\end{tabular}
\caption{The table shows the average and variance of the 
ratio between the total weight of the $k$-independent set obtained using GCN-greedy combo to that of the plain greedy algorithm for $k \in \{1,2,3,4\}$. We observe that the improvement is consistently 
above $5$ percent for the ER model and above $14$ percent 
for the BA model.}
\vspace{-1.5em}
\label{fig:table2}
\end{center}
\end{figure}

Interestingly, the GCN trained with ER graphs performs well on 
the BA data set as well. This indicates that the trained GCN
is transferable to other models.

\section{Conclusion}\label{section5}
In this paper, we investigated the well-studied
problem of link scheduling in 
wireless adhoc networks using the recent 
developments in graph neural networks.   
We modelled the wireless network as a  $k$-tolerant conflict graph and demonstrated that using a GCN, we can improve the 
performance of existing greedy algorithms. 
We have shown experimentally that this GCN 
model improves the performance of the greedy algorithm  
by at least $4$-$6$ percent for the ER model and 
$11$-$22$ percent for the BA model 
(depending on the value of $k$). 

In future, we would like to extend the model to a node 
dependent tolerance value $k_v$ and pass the tolerance value 
as the node features of the GNN in addition to the weights.

\bibliographystyle{ieeetr}
\bibliography{final_v1.bib}

\end{document}